\documentclass[%
 reprint,
superscriptaddress,
 amsmath,amssymb,
 aps,
 prl,
]{revtex4-2}
\usepackage[utf8]{inputenc}
\usepackage{graphicx}
\usepackage{diagbox}
\usepackage[normalem]{ulem}
\usepackage[colorlinks,bookmarks=false,linkcolor=blue,urlcolor=blue]{hyperref}
\usepackage[top=2.5cm,bottom=2.5cm,left=2cm,right=2cm]{geometry}
\usepackage{color}
\usepackage[dvipsnames]{xcolor}
\usepackage{wrapfig}
\usepackage{multirow}
\usepackage{mathtools}
\usepackage{gensymb}
\usepackage{graphicx}
\usepackage{float}
\usepackage{bm}
\usepackage{upgreek}
\usepackage{textcomp}
\usepackage{booktabs}
\usepackage{units}
\usepackage{setspace}
\usepackage{subfiles}
\usepackage{color, colortbl}
\usepackage{xcolor}
\usepackage{makecell}

\definecolor{mygray}{gray}{0.6}
\definecolor{LightCyan}{rgb}{0.88,1,1}
\definecolor{Mycolor2}{rgb}{1,0.88,1}

\def \be {\begin{equation}}
\def \ee {\end{equation}}

\DeclarePairedDelimiter\abs{\lvert}{\rvert}%
\DeclarePairedDelimiter\norm{\lVert}{\rVert}%

\makeatletter
\let\oldabs\abs
\def\abs{\@ifstar{\oldabs}{\oldabs*}}
\makeatother

\usepackage{titlesec}
\usepackage{lipsum}

\titleformat{\section}
  {\normalfont\scshape}{\thesection}{1em}{}

\date{\today}

\begin{document}

\title{Origin of correlated diffuse scattering in the hexagonal manganites}

\author{Tara N. Toši\'c}
 \email{tara.tosic@mat.ethz.ch}
 \affiliation{Materials Theory, ETH Zürich, Wolfgang-Pauli-Strasse 27, 8093 Zurich, Switzerland}

\author{Arkadiy Simonov}
 \affiliation{Disordered Materials, ETH Zürich, Vladimir-Prelog-Weg 1-5/10, 8093 Zürich, Switzerland}

 \author{Nicola A. Spaldin}
 \affiliation{Materials Theory, ETH Zürich, Wolfgang-Pauli-Strasse 27, 8093 Zurich, Switzerland}

%%%%%%%%%%%%%%%%%%%%%%%%%%%%%%%%%%%%%%%%%%%%%%%%%%%%%%%%%%%%%%%%%%%%%%%%%%%%%%%%%%%%%%%%%%%%%%%%%
%%%%%%%%%%%%%%%%%%%%%%%%%%%%%%%%%%%%%%%%ABSTRACT%%%%%%%5%%%%%%%%%%%%%%%%%%%%%%%%%%%%%%%%%%%%%%%%%
%%%%%%%%%%%%%%%%%%%%%%%%%%%%%%%%%%%%%%%%%%%%%%%%%%%%%%%%%%%%%%%%%%%%%%%%%%%%%%%%%%%%%%%%%%%%%%%%%
\begin{abstract} We use a combination of first-principles density functional calculations and spin-dynamics simulations to explain the unusual diffuse inelastic neutron scattering in the hexagonal multiferroic yttrium manganite, YMnO\textsubscript{3}. Using symmetry considerations, we construct a model spin Hamiltonian with parameters derived from our density functional calculations and show that it captures the measured behavior. We then show that the observed directionality in the structured diffuse scattering in momentum space is a hallmark of the triangular geometry, and that its persistence across a wide range of temperatures, both above and below the Néel temperature, T\textsubscript{N}, is a result of the strong magnetic frustration. We predict that this diffuse scattering exists in a yet-to-be-observed modulated continuum of energies, that its associated spin excitations have distinct in-plane and out-of-plane character and that the frustration influences the magnetism below the Néel temperature. Finally, we show that visualizing the magnetic order in terms of composite trimer magnetoelectric monopoles and toroidal moments, rather than individual spins, provides insight into the real space fluctuations, revealing clusters of emerging order in the paramagnetic state, as well as collective short-range excitations in the ordered antiferromagnetic phase.  Our understanding of this directional diffuse scattering in such a wide temperature range, both below and above T\textsubscript{N}, provides new insight into the magnetic phase transitions in frustrated systems. 
\end{abstract}
\maketitle
%%%%%%%%%%%%%%%%%%%%%%%%%%%%%%%%%%%%%%%%%%%%%%%%%%%%%%%%%%%%%%%%%%%%%%%%%%%%%%%%%%%%%%%%%%%%%%%%%
%%%%%%%%%%%%%%%%%%%%%%%%%%%%%%%%%%%%%%%%INTRODUCTION%%%%%%%%%%%%%%%%%%%%%%%%%%%%%%%%%%%%%%%%%%%%%%%%
%%%%%%%%%%%%%%%%%%%%%%%%%%%%%%%%%%%%%%%%%%%%%%%%%%%%%%%%%%%%%%%%%%%%%%%%%%%%%%%%%%%%%%%%%%%%%%%%%
Magnetic diffuse scattering, a measure of local magnetic correlation, is ubiquitous in finite-temperature inelastic neutron scattering, where it reveals complex underlying physics \cite{mirebeaua2017diffuse}. It can be particularly strong in magnetically frustrated systems, where competing magnetic interactions give rise to magnetic fluctuations, suppressing the long-range magnetic order to lower temperatures than their Curie-Weiss temperature suggests \cite{khomskii2014_exchange}. In such systems, short-range and long-range correlations can co-exist over large temperature regimes, giving rise to exotic magnetic states, such as spin liquids \cite{balents2010spin}, spin ice \cite{bramwell2001spinice, drisko2010topological} or spin glasses \cite{anderson1978theconcept}, and even influencing superconductivity \cite{glasbrenner2015effect,gu2022frustrated}. In these cases, diffuse scattering can provide useful insight into non collinear magnetic structures \cite{long1996magnetic} and the nature of phase transitions \cite{copley1995orientational}.\\
%%%%%%%%%%%%%%%%%%%%%%%%%%%%%%%%%%%%%%%%%%%%%%%%%%%%%%%%%%%%%%%%%%%%%%%%%%%%%%%%%%%%%%%%%%%%%%%%%
%%%%%%%%%%%%%%%%%%%%%%%%%%%%%%%%%%%%%%%%%%%%%%%%%%%%%%%%%%%%%%%%%%%%%%%%%%%%%%%%%%%%%%%%%%%%%%%%%
%%%%%%%%%%%%%%%%%%%%%%%%%%%%%%%%%%%%%%%%%%%%%%%%%%%%%%%%%%%%%%%%%%%%%%%%%%%%%%%%%%%%%%%%%%%%%%%%%
\indent The hexagonal manganites, of which h-YMnO\textsubscript{3} is the prototype, provide a natural playground for studying triangular magnetic frustration, since they contain triangular $ab$ planes of antiferromagnetically coupled Mn ions that are only weakly coupled along the $c$ direction \cite{janas2021classical,sato2003unconventional,oh2016spontaneous,das2014bulk}.
Inelastic neutron scattering has revealed unconventional scattering signals in this system, such as low-energy sub-T\textsubscript{N} quasi-elastic continua \cite{sato2003unconventional}, inelastic peaks above T\textsubscript{N} \cite{chatterji2008neutron}, and diffuse scattering well above and below T\textsubscript{N},  \cite{park2003magnetic} as well as signatures of magnon decay, such as roton-like minima, flat modes and line-width broadening \cite{oh2013magnonbreakdown,oh2016spontaneous}. These behaviors have been variously attributed to closely related phenomena such as cooperative paramagnetism \cite{sato2003unconventional}, Kosterlitz-Thouless phases \cite{sato2003unconventional} and classical spin liquids \cite{janas2021classical}, or attributed to quasi-particle breakdown \cite{oh2016spontaneous,oh2013magnonbreakdown}. Our paper focuses primarily on the persistence of directional ``rod-like" diffuse magnetic excitations which have been observed to persist across T\textsubscript{N} in an unusually large temperature regime \cite{janas2021classical}, and whose origin is not understood. We simulate neutron scattering results from Ref.~\cite{janas2021classical} using a combined first-principles and spin-dynamics approach and clarify the nature of the short-range correlations, as well as their impact on the low-temperature magnetism in h-YMnO\textsubscript{3}. \\
%%%%%%%%%%%%%%%%%%%%%%%%%%%%%%%%%%%%%%%%%%%%%%%%%%%%%%%%%%%%%%%%%%%%%%%%%%%%%%%%%%%%%%%%%%%%%%%%%
%%%%%%%%%%%%%%%%%%%%%%%%%%%%%%%%%%%%%%%%%%%%%%%%%%%%%%%%%%%%%%%%%%%%%%%%%%%%%%%%%%%%%%%%%%%%%%%%%
%%%%%%%%%%%%%%%%%%%%%%%%%%%%%%%%%%%%%%%%%%%%%%%%%%%%%%%%%%%%%%%%%%%%%%%%%%%%%%%%%%%%%%%%%%%%%%%%%
\begin{figure}
    \centering
    \includegraphics[width=1\linewidth]{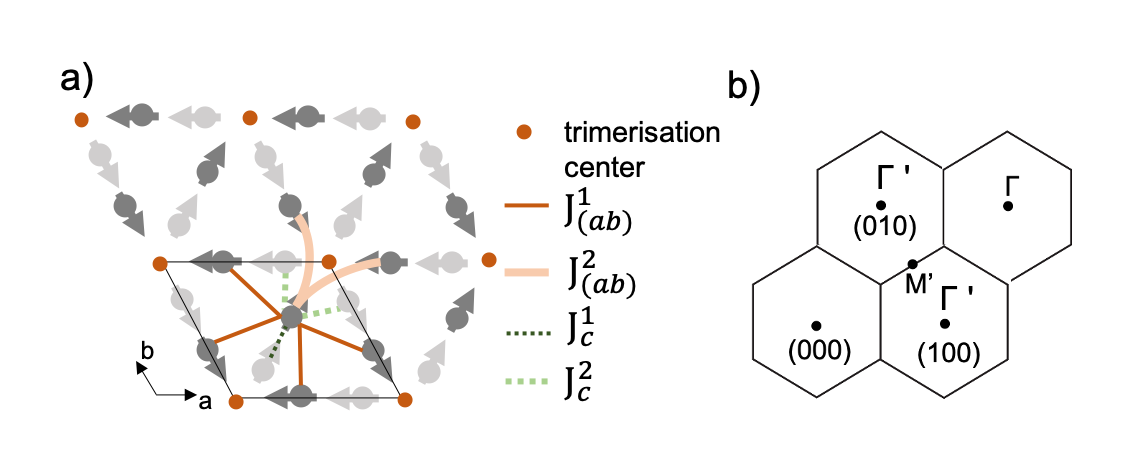}
    \caption{a) Magnetic ground state of h-YMnO\textsubscript{3} and symmetric exchanges of one representative spin. Lighter and darker colored arrows represent Mn spins belonging to consecutive $ab$ planes along $c$ and the black lines indicate the boundaries of the six Mn-atom magnetic unit cell. Red dots indicate trimerisation sites. b) Points of interest in reciprocal space. Black lines indicate the reciprocal lattice and the magnetic and structural Bragg peaks occur at $\Gamma '$ and $\Gamma$, respectively.}
    \label{fig:ham_kpath}
\end{figure}
\renewcommand{\arraystretch}{2}
%%%%%%%%%%%%%%%%%%%%%%%%%%%%%%%%%%%%%%%%%%%%%%%%%%%%%%%%%%%%%%%%%%%%%%%%%%%%%%%%%%%%%%%%%%%%%%%%%
%%%%%%%%%%%%%%%%%%%%%%%%%%%%%%%%%%%%%%%%%%%%%%%%%%%%%%%%%%%%%%%%%%%%%%%%%%%%%%%%%%%%%%%%%%%%%%%%%
%%%%%%%%%%%%%%%%%%%%%%%%%%%%%%%%%%%%%%%%%%%%%%%%%%%%%%%%%%%%%%%%%%%%%%%%%%%%%%%%%%%%%%%%%%%%%%%%%
\indent Within each h-YMnO\textsubscript{3} $ab$ plane, the S=2 Mn\textsuperscript{3+} lattice distinguishes itself from a perfect triangular frustrated lattice (TFL) through a trimerisation which displaces the Mn ions away from their ideal 1/3 position \cite{tosic2022influence}. Within the trimerised structure, MnO\textsubscript{5} bipyramids tilt towards their corner sharing oxygen, and layers of yttrium sites, which separate the magnetic planes, buckle \cite{howard2013crystal}. The outcome of this particular crystal chemistry is a hierarchy of magnetic interactions, which are important to define the system's magnetic ground state \cite{lorenz2013hexagonal,das2014bulk,tosic2022influence}. Following our previous work in Ref.~\cite{tosic2022influence}, we describe the magnetism using the following Hamiltonian for unit vector magnetic moments $\hat{e}_i$, the parameters of which are listed in Table \ref{tab:ham_parameters}:
\begin{align}\label{Hamiltonian}
    \mathcal{H} = & J_{1}^{ab}\sum_{\substack{\left<i,j\right>\\\text{1\textsuperscript{st}n.n}}}\hat{e}_i\cdot\hat{e}_j
   + J_{2} ^{ab}\sum_{\substack{\left<i,j\right>\\\text{2\textsuperscript{nd}n.n}}}\hat{e}_i\cdot\hat{e}_j \\
  \nonumber&  + J_1^z\sum_{\substack{\left<i,j\right>\\\text{3\textsuperscript{rd}n.n}}}\hat{e}_i\cdot\hat{e}_j  
   + J_2^z\sum_{\substack{\left<i,j\right>\\\text{4\textsuperscript{th}n.n}}}\hat{e}_i\cdot\hat{e}_j \\
    \nonumber&             +\sum_i A_{zz}(e_{i,z})^2+\sum_iA_{x^*x^*}(e_{i,x^*})^2 \quad.
\end{align}
Here, the first and second lines list the symmetric exchange interactions, with the in-plane dominant first and second nearest neighbor AFM interactions, $J_{1}^{ab}$ and $J_{2}^{ab}$ respectively, and the weak inter-planar ferromagnetic (FM) interactions, $J_1^z$ and $J_2^z$ (see Fig.~\ref{fig:ham_kpath}a)). The third line describes the single ion anistropy (SIA), which we compute to have a hard axis along $c$, described by $A_{zz}$ and a weaker in-plane anisotropy, where $x^*$ denotes the local in-plane easy-axis direction, which lies along the vector connecting the magnetic site to its trimerisation center (see Fig.~\ref{fig:ham_kpath}a)).\\
\begin{table}
    \centering
    \begin{tabular}{ccccccc}
    \toprule
        & $J_{1}^{ab}$ & $J_{2} ^{ab}$ & $J_1^z$  & $J_2^z$  & $ A_{zz}$  & $A_{x^*x^*}$ \\
    \hline
    DFT \cite{tosic2022influence}  & 7.779 & 4.874 & -0.015 & -0.003  &0.530  & -0.117 \\
    \bottomrule
    \end{tabular}
    \caption{Parameters extracted from density functional theory (DFT) in units of meV.}
    \label{tab:ham_parameters}
\end{table}
\begin{figure}
    \centering
    \includegraphics[width=1\linewidth]{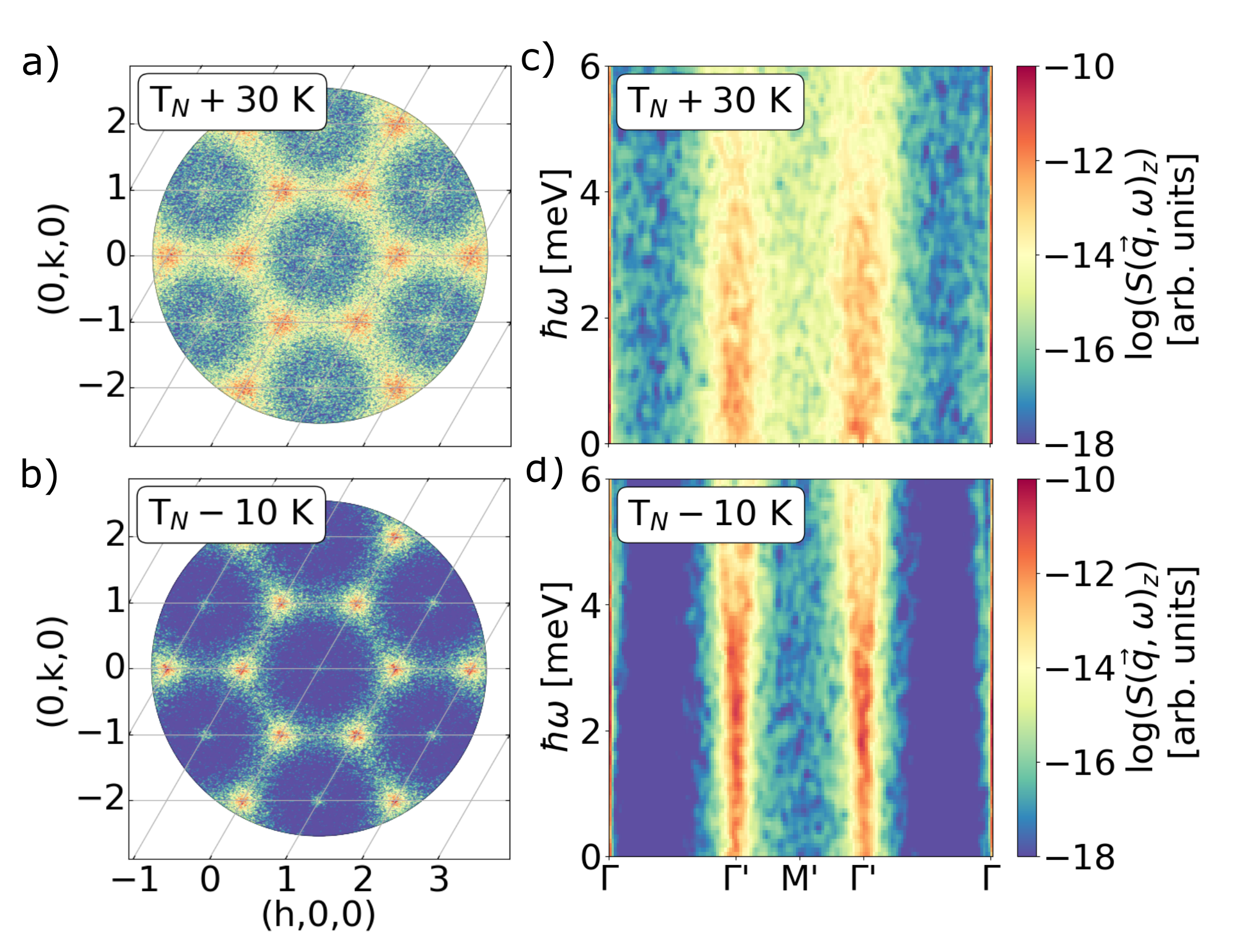}
    \caption{Spin-dynamics simulated dynamical structure factor, above (top panels) and below (bottom panels) T\textsubscript{N}. a)-b) Constant $\hbar\omega=1.04$\,meV cut. c)-d) Calculated low energy spectrum along the $\Gamma$-$\Gamma'$-$\text{M}'$-$\Gamma'$-$\Gamma$ path, sketched in Fig.~\ref{fig:ham_kpath}b).}
    \label{fig:prl_res}
\end{figure}
%%%%%%%%%%%%%%%%%%%%%%%%%%%%%%%%%%%%%%%%%%%%%%%%%%%%%%%%%%%%%%%%%%%%%%%%%%%%%%%%%%%%%%%%%%%%%%%%%
%%%%%%%%%%%%%%%%%%%%%%%%%%%%%%%%%%%COMPUTATIONAL DETAILS%%%%%%%%%%%%%%%%%%%%%%%%%%%%%%%%%%%%%%%%%%%%%%%
%%%%%%%%%%%%%%%%%%%%%%%%%%%%%%%%%%%%%%%%%%%%%%%%%%%%%%%%%%%%%%%%%%%%%%%%%%%%%%%%%%%%%%%%%%%%%%%%%
\indent We use the parameters from Table~\ref{tab:ham_parameters} in our Hamiltonian in Eq.~\ref{Hamiltonian} to calculate the spin dynamics using the UppASD code \cite{antropov1995abinitio,skubic2008method,etz2008atomistic,eriksson2017abinition}. We note that while density functional theory (DFT) over-estimates interaction parameters, their relative strengths are comparable to those fitted to inelastic neutron scattering data (which we list in Table S1 in the Supplementary Information): $\frac{|J^z|}{J^{ab}}\simeq0.002$ and $\frac{A_{zz}}{J^{ab}}\simeq 0.09$ \cite{tosic2022influence}, where $J^z$ and $J^{ab}$ are the average intra- and inter-planar exchanges, respectively. We compute energy dispersions by converting spin-spin correlations into the usual dynamical structure factor $S(\vec{q},\omega)$, as implemented in UppASD, using this quantity to compare our theoretical model predictions with experimental scattering data \cite{sturm1993dynamic}. We create a $50$$\times50$$\times10$ supercell, cool our system from 400\,K down to 2\,K, going in steps of 2\,K as the temperature drops below 100\,K, evolving the system at each temperature for 60000 femtosecond time-steps; this yields an energy resolution of at least $0.07$\,meV, which matches the experimental resolution of Ref.~\cite{janas2021classical}. After initializing the spin directions randomly and cooling the system down to T\textsubscript{N}-50\,K, the system stabilizes in its ground state configuration (shown in Fig.~\ref{Hamiltonian}a)), indicating that the Hamiltonian of Eq.~\ref{Hamiltonian} with the parameters of Table~\ref{tab:ham_parameters} correctly captures the experimentally observed ground state \cite{fiebig2000determination}.\\
\begin{figure*}
    \centering
    \includegraphics[width=1\linewidth]{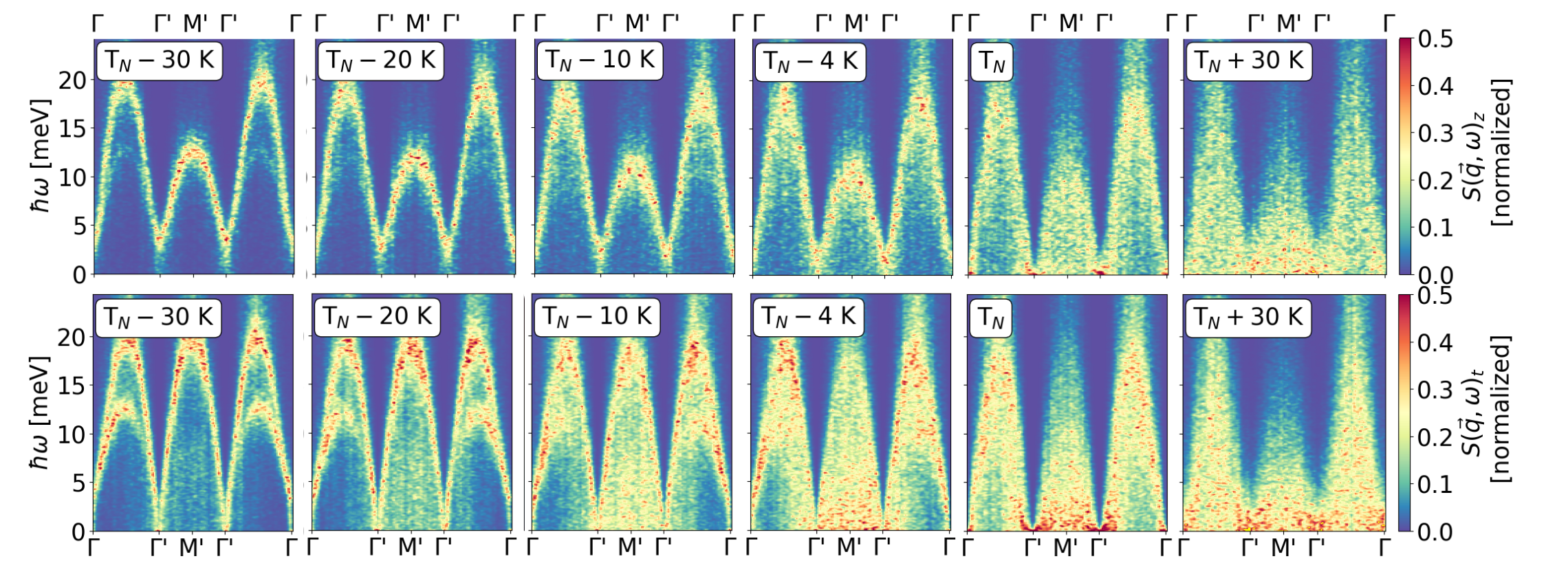}
    \caption{Calculated temperature dependence of the out-of-plane $z$ (top panel) and transverse $t$ (bottom panel) components of the dynamical structure factor, $S(\vec{q},\omega)_z$ and $S(\vec{q},\omega)_t$, respectively, along the $\Gamma$-$\Gamma'$-$\text{M}'$-$\Gamma'$-$\Gamma$ path.}
    \label{fig:spinon}
\end{figure*}
%%%%%%%%%%%%%%%%%%%%%%%%%%%%%%%%%%%%%%%%%%%%%%%%%%%%%%%%%%%%%%%%%%%%%%%%%%%%%%%%%%%%%%%%%%%%%%%%%
%%%%%%%%%%%%%%%%%%%%%%%%%%%%%%%%%%%%%%%%%%%%%RESULTS%%%%%%%%%%%%%%%%%%%%%%%%%%%%%%%%%%%%%%%%%%%%%%%
%%%%%%%%%%%%%%%%%%%%%%%%%%%%%%%%%%%%%%%%%%%%%%%%%%%%%%%%%%%%%%%%%%%%%%%%%%%%%%%%%%%%%%%%%%%%%%%%%
\indent We now present our calculated dynamical structure factors and compare them to the inelastic neutron scattering measurements from Ref.~\cite{janas2021classical}. Since Ref.~\cite{janas2021classical} only measures the component of the spins perpendicular to the momentum transfer, we first plot our calculated $S(\vec{q},\omega)_z$ in the $(hk0)$ plane in Fig. \ref{fig:prl_res} at 30\,K above and 10\,K below T\textsubscript{N}. These should be compared with the measured spectra in Fig.1 in Ref. \cite{janas2021classical}.  Figs. \ref{fig:prl_res}\,a) and b) are in-plane $(hk0)$ cuts at constant $\hbar\omega=1.04$\,meV. Our simulations reproduce the rod-like diffuse scattering observed in Ref. \cite{janas2021classical} well above and below T\textsubscript{N}, connecting neighboring magnetic Bragg peaks ($\Gamma'$) along the $(h00)$, $(0k0)$ and $(hh0)$ directions.  We reproduce other characteristics of the experimental data, namely the narrowing of the diffuse rod width in $\vec{q}$-space, as well as the decrease in diffuse scattering around the $\Gamma'$ points with decreasing temperature. In Figs.~\ref{fig:prl_res}\,c) and d), we plot the frequency dependence of $S(\vec{q},\omega)_z$ along the $\Gamma$-$\Gamma'$-$\text{M}$-$\Gamma'$-$\Gamma$ $\vec{q}$-path (see Fig.~\ref{fig:ham_kpath}b)). Our simulations recreate the measured sheet of diffuse scattering along the $\Gamma'$-$\text{M}$-$\Gamma'$ rod path, as well as its attenuation with decreasing temperature. Below T\textsubscript{N}, a $\Gamma'$ gapped excitation appears as a decrease in low frequency intensity and a condensation of the spectrum at around $\Delta=2$\,meV, shown in Fig.~\ref{fig:prl_res}c). The amplitude of this excitation gap is in reasonable agreement with the experimentally measured gap $\Delta\simeq1.7$\,meV in Ref~\cite{holm2018magneticphase,janas2021classical}.\\
%%%%%%%%%%%%%%%%%%%%%%%%%%%%%%%%%%%%%%%%%%%%%%%%%%%%%%%%%%%%%%%%%%%%%%%%%%%%%%%%%%%%%%%%%%%%%%%%%
%%%%%%%%%%%%%%%%%%%%%%%%%%%%%%%%%%%%%%%%%%%%%%%%%%%%%%%%%%%%%%%%%%%%%%%%%%%%%%%%%%%%%%%%%%%%%%%%%
%%%%%%%%%%%%%%%%%%%%%%%%%%%%%%%%%%%%%%%%%%%%%%%%%%%%%%%%%%%%%%%%%%%%%%%%%%%%%%%%%%%%%%%%%%%%%%%%%
\indent We now determine which terms in our Hamiltonian explain the behavior of Fig.~\ref{fig:prl_res} by setting individual parameters to zero, and comparing the resulting calculated inelastic magnetic spectra. We find that diffuse rods persist even a simple model (see Fig. S1 in Supplementary Information) in which we only include the first-nearest neighbor AFM symmetric exchanges. This shows that strong geometric frustration arising from AFM-coupled classical spins on a triangular network causes the correlated rod-like diffuse scattering. In our simulations, anisotropic short-ranged correlated fluctuations persist above and below T\textsubscript{N}, and we capture their temperature dependence. Note that Ref~\cite{janas2021classical} considers scattering arising from clusters of coherently oscillating spins, and introduces an exponential decay in the spin-spin correlation length. Since their Hamiltonian is, by definition, isotropic, it cannot describe the $\vec{q}$-space directionality of the short-range correlations. Additionally, we find that the low-temperature energy dispersion is affected by $A_{zz}$, with the 2\,meV $\Gamma'$ gapped excitations (shown in Fig.~\ref{fig:prl_res}c)) disappearing when we turn off the hard axis SIA (see Fig~S2 in the Supplementary Information).\\
\begin{figure*}
    \centering\includegraphics[width=0.95\linewidth]{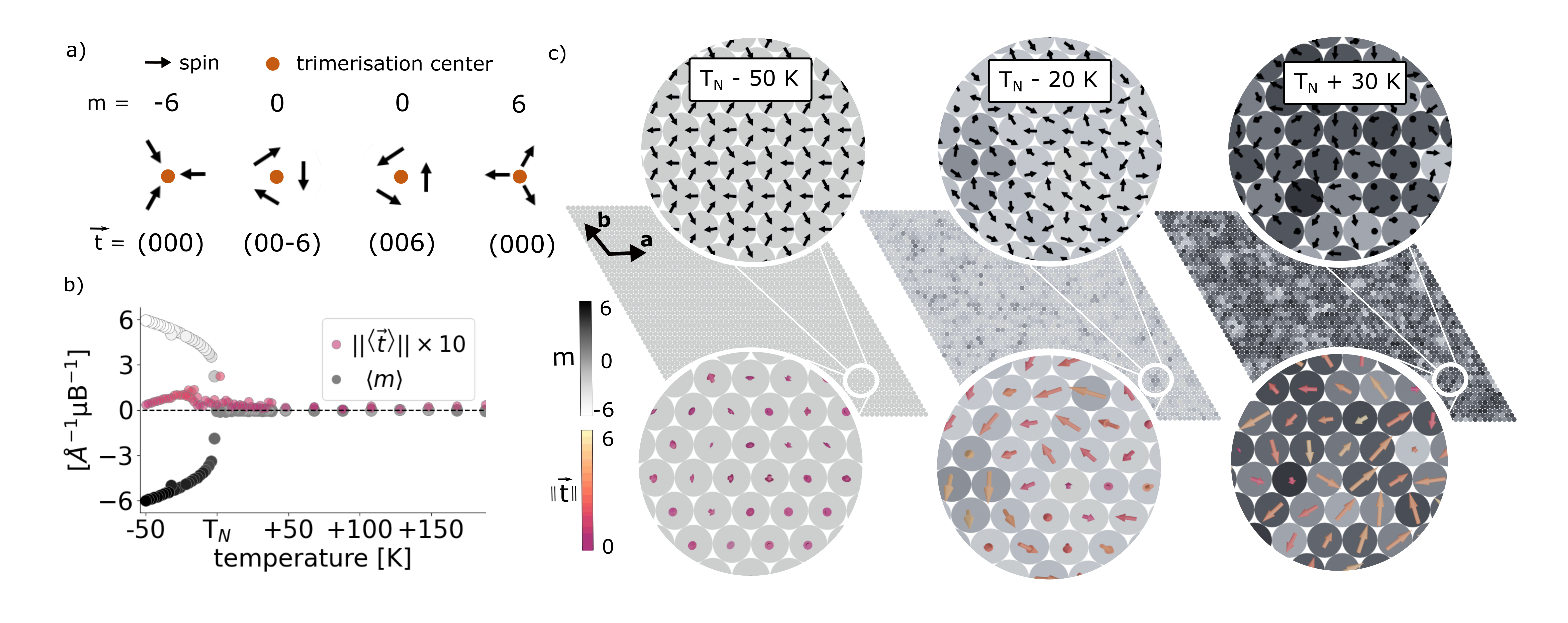}
    \caption{Real space visualization of the spins. a) Planar monopoles and toroids. The fractional coordinates within each trimer of spins are, clock-wise from the top left spin, $(\frac{1}{3}+\delta)$($-\vec{a}+\vec{b}$), $(\frac{1}{3}+\delta)$$\vec{a}$ and $(\frac{1}{3}+\delta)$($-\vec{a}-\vec{b}$), where $\delta$ is a small Mn off-centering. b) Temperature evolution of the trimer monopolar expectation value $\left<m\right>$ of one simulation, averaged over layers converging to a positive $m=+6$ (white) and negative $m=-6$ (black) monopolar ground state, as well as of ten times the norm of the average trimer toroidal moment. c) Spin configurations (top insets) and their associated trimer monopoles (gray-scale disks) and toroids (colored vectors in the bottom insets) at three different temperatures.}
    \label{fig:OPvsT}
\end{figure*}
%%%%%%%%%%%%%%%%%%%%%%%%%%%%%%%%%%%%%%%%%%%%%%%%%%%%%%%%%%%%%%%%%%%%%%%%%%%%%%%%%%%%%%%%%%%%%%%%%
%%%%%%%%%%%%%%%%%%%%%%%%%%%%%%%%%%%%%%%%%%%%%%%%%%%%%%%%%%%%%%%%%%%%%%%%%%%%%%%%%%%%%%%%%%%%%%%%%
%%%%%%%%%%%%%%%%%%%%%%%%%%%%%%%%%%%%%%%%%%%%%%%%%%%%%%%%%%%%%%%%%%%%%%%%%%%%%%%%%%%%%%%%%%%%%%%%%
\indent In the following, we analyze the high frequency and spin-component dependence of the rod-like scattering, and predict and explain additional characteristics of the diffuse signal, that have not yet been measured experimentally.\\ 
\indent We begin by describing the calculated low temperature magnon branches in the top left plot of Fig.~\ref{fig:spinon}) for $S(\vec{q},\omega)_z$, where we normalize along each $\vec{q}$ point to showcase the fainter higher energy and diffuse scattering (note that our energy dispersions match earlier non-linear spin wave theory calculations \cite{petit2007spinphonon,holm2018magnetic}). We distinguish between spectra along paths linking two magnetic Bragg peaks, along the $\Gamma'$-M-$\Gamma'$ rod directions, and paths along $\Gamma$-$\Gamma'$ ($\equiv\Gamma'$-$\Gamma$) that join a structural and a magnetic Bragg peak. Along $\Gamma$-$\Gamma'$, we see a low- and a high-energy magnon branch, peaking at around 13 and 20\,meV, respectively, and only the former visibly appears along $\Gamma'$-M$'$-$\Gamma'$.\\
%%%%%%%%%%%%%%%%%%%%%%%%%%%%%%%%%%%%%%%%%%%%%%%%%%%%%%%%%%%%%%%%%%%%%%%%%%%%%%%%%%%%%%%%%%%%%%%%%
%%%%%%%%%%%%%%%%%%%%%%%%%%%%%%%%%%%%%%%%%%%%%%%%%%%%%%%%%%%%%%%%%%%%%%%%%%%%%%%%%%%%%%%%%%%%%%%%%
%%%%%%%%%%%%%%%%%%%%%%%%%%%%%%%%%%%%%%%%%%%%%%%%%%%%%%%%%%%%%%%%%%%%%%%%%%%%%%%%%%%%%%%%%%%%%%%%%
\indent Next, we focus on the temperature dependence of the $S(\vec{q},\omega)_z$ signal, shown in the top panel of Fig.~\ref{fig:spinon}, at five different temperatures. The first notable feature are the tails of diffuse scattering above the previously described magnon branches. However, most of the scattering occurs below the magnon branches: in particular along $\Gamma'$-M$'$-$\Gamma'$, a remarkable continuum of excitations, bounded at high frequency by the magnon branch, starts to appear at 4\,K below T\textsubscript{N}, and persists up to 150\,K above T\textsubscript{N} (see higher temperature calculations in Figs.~S3 and S4 in Supplementary Information). Fluctuations at frequencies up to that of the magnon branch energy therefore dominate the diffuse spectrum. Similar excitation continua occur along $\Gamma$-$\Gamma '$, but have, in addition to an upper bound, a lower bound below T\textsubscript{N}, set by the lower magnon branch.\\
%%%%%%%%%%%%%%%%%%%%%%%%%%%%%%%%%%%%%%%%%%%%%%%%%%%%%%%%%%%%%%%%%%%%%%%%%%%%%%%%%%%%%%%%%%%%%%%%%
%%%%%%%%%%%%%%%%%%%%%%%%%%%%%%%%%%%%%%%%%%%%%%%%%%%%%%%%%%%%%%%%%%%%%%%%%%%%%%%%%%%%%%%%%%%%%%%%%
%%%%%%%%%%%%%%%%%%%%%%%%%%%%%%%%%%%%%%%%%%%%%%%%%%%%%%%%%%%%%%%%%%%%%%%%%%%%%%%%%%%%%%%%%%%%%%%%%
\indent Next, we calculate the as-yet-unmeasured transverse component $S(\vec{q},\omega)_t$, shown in the bottom panel of Fig.~\ref{fig:spinon}, to compare with $S(\vec{q},\omega)_z$ presented above. Comparing the top and bottom panels of Fig.~\ref{fig:spinon}, we see no significant differences between the in-plane and out-of-plane dynamical structure factors above T\textsubscript{N} (see the T\textsubscript{N}+30\,K plot in Fig.~\ref{fig:spinon}, as well as the higher temperature plots in Figs.~S3 and S4 in Supplementary Information). Notable differences occur below T\textsubscript{N} however, where we see that the bounded $S(\vec{q},\omega)_t$ excitation continuum is more clearly defined than that of $S(\vec{q},\omega)_z$ along the whole $\Gamma$-$\Gamma'$-M$'$-$\Gamma'$-$\Gamma$ path, and that, along $\Gamma'$-M$'$-$\Gamma'$, it is now the upper-magnon branch (as oppose to the lower-magnon branch for $S(\vec{q},\omega)_z$) modulating the higher-energy scattering. These bounded regions of continuous excitations are a classical analog to the multi-spinon continua observed in low dimensional magnetically frustrated quantum systems \cite{lake2005elastic,zheng2006anomalous,ghioldi2015magnons}.\\
%%%%%%%%%%%%%%%%%%%%%%%%%%%%%%%%%%%%%%%%%%%%%%%%%%%%%%%%%%%%%%%%%%%%%%%%%%%%%%%%%%%%%%%%%%%%%%%%%
%%%%%%%%%%%%%%%%%%%%%%%%%%%%%%%%%%%%%%%%%%%%%%%%%%%%%%%%%%%%%%%%%%%%%%%%%%%%%%%%%%%%%%%%%%%%%%%%%
%%%%%%%%%%%%%%%%%%%%%%%%%%%%%%%%%%%%%%%%%%%%%%%%%%%%%%%%%%%%%%%%%%%%%%%%%%%%%%%%%%%%%%%%%%%%%%%%%
\indent Again, we rationalize the spectra by setting parameters in our Hamiltonian to zero and re-calculating $S(\vec{q},\omega)_{t,z}$. Our main finding is that the hard axis SIA plays a defining role in distinguishing  between the $S(\vec{q},\omega)_{t}$ and $S(\vec{q},\omega)_{z}$ spectral weight distributions in the low temperature diffuse spectrum. When $A_{zz}=0$, there is no preference for the spins to lie in the $ab$ plane, and both $S(\vec{q},\omega)_{t}$ and $S(\vec{q},\omega)_{z}$ carry both the upper and lower magnon branches along $\Gamma'$-M$'$-$\Gamma'$ (Fig. S5 in the Supplementary Information). When the spins adopt the in-plane configuration favored by the SIA, in-plane and out-of-plane fluctuations become inequivalent. Since short-range out-of-plane fluctuations can be accommodated without breaking the local 120$\degree$ nearest-neighbor antiferromagnetic configuration, they contribute to the low energy mode that emerges in the out-of-plane spectrum below T\textsubscript{N} along the $\Gamma'$-$\text{M}'$-$\Gamma'$ rod path, seen in the top panel of Fig.~\ref{fig:spinon}. On the other hand, in-plane fluctuations break local 120$\degree$ nearest-neighbor order and so occur at higher frequency, with dominant spectral weight around the upper magnon branch (lower panel of Fig.~\ref{fig:spinon}). \\
\indent Summarizing our understanding of our simulated spectra so far, from at least 150\,K above T\textsubscript{N} down to T\textsubscript{N}, the SIA is far smaller than the thermal energy: rod-like diffuse scattering originates from 1st nearest neighbor AFM interactions between all three $x,y$ and $z$ spin components, and we can neglect the SIA and the weak inter-plane coupling. On further cooling through T\textsubscript{N}, we observe the effect of the SIA which breaks the local spin symmetry and the in-plane and out-of-plane correlations start to differ.\\ 
%%%%%%%%%%%%%%%%%%%%%%%%%%%%%%%%%%%%%%%%%%%%%%%%%%%%%%%%%%%%%%%%%%%%%%%%%%%%%%%%%%%%%%%%%%%%%%%%%
%%%%%%%%%%%%%%%%%%%%%%%%%%%%%%%%%%%%%%%%%%%%%%%%%%%%%%%%%%%%%%%%%%%%%%%%%%%%%%%%%%%%%%%%%%%%%%%%%
%%%%%%%%%%%%%%%%%%%%%%%%%%%%%%%%%%%%%%%%%%%%%%%%%%%%%%%%%%%%%%%%%%%%%%%%%%%%%%%%%%%%%%%%%%%%%%%%%
\indent Next, we visualize the correlations in real space. Since triangular arrangements of spins dominate the spectrum in the Néel phase, we find it useful to visualize excitations away from the ground state by associating two multipolar quantities, the local magnetoelectric monopole $m$ and toroidal moment $\vec{t}$ \cite{spaldintoroidal2008} of the trimer, with each trimer of spins belonging to the same trimerisation center (hereafter we refer to these as a monopole and toroid, respectively):
\begin{align}
    m&=\sum_{i}\vec{r}_i\cdot{\hat{e}_i} \quad ,\\
    \vec{t}&=\sum_{i}\vec{r_i}\times{\hat{e}_i} \quad .
\end{align}
Here the sum is over the three spin sites, $\vec{r}_i$ is the position of spin $i$ in fractional lattice coordinates, relative to the trimer center, and $\hat{e}_i$ is the unit vector describing the direction of the spin. The h-YMnO\textsubscript{3} ground state (shown in Fig.\ref{fig:ham_kpath}a)) is described by an antiferromonopolar order, corresponding to $m=\pm6.1\simeq\pm6$ monopoles (and corresponding $\vec{t}=(000)$) alternating along $c$ (see Fig.~\ref{fig:OPvsT}a).\\ 
\indent In Fig.~\ref{fig:OPvsT}b), as a function of temperature, we plot the expectation value of $m$, $\left<m\right>$,  for layers which reach a $+6$ (white) or $-6$ (black) monopolar ground state at low temperature, as well as the norm of the system's average $\vec{t}$ vector, $\norm{\left<\vec{t}\right>}$. At 0\,K,  $\left<m\right>$ and $\norm{\left<\vec{t}\right>}$ converge to their ground state values, $\pm6$ and 0 respectively; we illustrate a close-to perfect $m=-6$ monopolar arrangement of spins in a single $ab$ layer at T\textsubscript{N}$-50$\,K in Fig.~\ref{fig:OPvsT}c). Collective excitations around this ground state give rise to the low temperature $S(q,\omega)_{t,z}$ magnon branches in Fig.~\ref{fig:spinon}. As the system approaches T\textsubscript{N} from below, long-range order gradually gives way to short-range fluctuations away from the monopolar ground state: we see the expected decrease in magnitude in Fig.~\ref{fig:OPvsT}b) in both $\left<m\right>$ branches, and a small increase in $\norm{\left<\vec{t}\right>}$. The real-space behavior in this temperature regime is illustrated by the multipolar projections at T\textsubscript{N}$-20$\,K in Fig.~\ref{fig:OPvsT}b), where we observe perturbations of the order in the monopolar projection, and a corresponding short-range swirl-like pattern in the $\vec{t}$ inset. These correlated spatial fluctuations coincide with the increase in the correlated diffuse scattering intensity, coexisting with the spin-wave dispersion in Fig.~\ref{fig:spinon}. As temperature increases above T\textsubscript{N}, $\left<m\right>$ and $\norm{\left<\vec{t}\right>}$ fluctuate around 0, ultimately converging to 0. We show the spin configuration at T\textsubscript{N}+30\,K in Fig.~\ref{fig:OPvsT}c), where we see nanoscale regions of similar monopolar value indicating that spins still have a preferred local 120$\degree$ nearest-neighbor configuration. We observe in the $\vec{t}$ correlation length a significant decrease, compared to its value below T\textsubscript{N}, consistent with the calculated increase in dispersion width with increasing temperature in Fig.~\ref{fig:spinon}.\\
%%%%%%%%%%%%%%%%%%%%%%%%%%%%%%%%%%%%%%%%%%%%%%%%%%%%%%%%%%%%%%%%%%%%%%%%%%%%%%%%%%%%%%%%%%%%%%%%%%%%%%%%%%%%%
%%%%%%%%%%%%%%%%%%%%%%%%%%%%%%%%%%%%%%%%%%%%%%%%%SUMMARY & OUTLOOK%%%%%%%%%%%%%%%%%%%%%%%%%%%%%%%%%%%%%%%%%%%%%%%
%%%%%%%%%%%%%%%%%%%%%%%%%%%%%%%%%%%%%%%%%%%%%%%%%%%%%%%%%%%%%%%%%%%%%%%%%%%%%%%%%%%%%%%%%%%%%%%%%%%%%%%%%%%%%
\indent In summary, we have reproduced structured diffuse inelastic neutron scattering from first-principles, capturing its $\vec{q}$-space directionality and frequency dependence using a combination of a DFT-extracted magnetic Hamiltonian and spin dynamics. New features of this diffuse scattering are also predicted, namely its transverse character, the strong non-linearity of its associated spin-spin correlations, its bounded excitation continuum as well as possible additional regions in $\vec{q}$-space where similar diffuse scattering could exist. We provide intuition on the spin excitations responsible for the scattering through a trimer multipolar formalism, which helps us visualize short-range correlations beyond the 120\degree spin arrangement, highlighting the surprisingly large out-of-plane correlated spatial fluctuations in the Néel state. We also show that the planar magnetoelectric monopole $\left<m\right>$ is a good order parameter for this system. Our work motivates experimental measurements of the continuous scattering, $t$ and $z$-component resolved structure factors and diffuse rod reciprocal space direction in $h$-YMnO\textsubscript{3}. We hope that it inspires theoretical characterization of such structured diffuse scattering in other systems.\\
\indent{\textit{Acknowledgments.}} We thank Johan Hellsvik and Anders Bergmann for their invaluable help with the UppASD code framework, as well as Kim Lefmann, Sophie Holm-Janas and Jian-Rui Soh for the fruitful discussions. This work was funded by ETH Zürich and the European Research Council (ERC) under the European Union’s Horizon 2020 research and innovation program project HERO grant agreement No. 810451. Computational resources were provided by ETH Zürich and the Swiss National Supercomputing center, project IDs eth3 and s889.

\bibliography{ref.bib}

\end{document}

\onecolumngrid

\title{Supplementary Information}

\author{Tara N. Tošić}
 \email{tara.tosic@mat.ethz.ch}
 \affiliation{Materials Theory, ETH Zürich, Wolfgang-Pauli-Strasse 27, 8093 Zurich, Switzerland}
 
\author{Arkadiy Simonov}
 \affiliation{Disordered Materials, ETH Zürich, Vladimir-Prelog-Weg 1-5/10, 8093 Zürich, Switzerland}
 \author{Nicola A. Spaldin}
 \affiliation{Materials Theory, ETH Zürich, Wolfgang-Pauli-Strasse 27, 8093 Zurich, Switzerland}

\maketitle

\onecolumngrid
\newpage

%%%%%%%%%%%%%%%%%%%%%%%%%%%%%%%%%%%%%%%%%%%%%%%%%%%%%%%%%%%%%%%%%%%%%%%%%%%%%%%%%%%%%%%%%%%%%%%%%
%%%%%%%%%%%%%%%%%%%%%%%%%%%%%%%%%%%%SUPPLEMENTARY INFORMATION%%%%%%%%%%%%%%%%%%%%%%%%%%%%%%%%%%%%%%%%
%%%%%%%%%%%%%%%%%%%%%%%%%%%%%%%%%%%%%%%%%%%%%%%%%%%%%%%%%%%%%%%%%%%%%%%%%%%%%%%%%%%%%%%%%%%%%%%%%%%

\begin{table}[H]
    \centering
    \begin{tabular}{ccccccc}
    \toprule
    \toprule
         parameter & $J_{1}^{(ab)}$ & $J_{2} ^{(ab)}$ & $J_{1}^c$  & $J_{2}^c$  & $ A_{zz}$  & $A_{x^*x^*}$ \\
    \hline
    \cite{janas2021classical} &2.4  & / & 0.144 & / &0.312  & / \\
    \cite{sato2003unconventional}  & 3.4  &2.02  & -0.014$^{\dagger}$ &0  & 0.28  & -0.0007(6)\\
    \cite{oh2016spontaneous} & 4  & 1.8 & / & / & 0.28 & -0.02\\
        \cite{chatterji2008neutron}  & 2.4 & / & / & / & 0.31 & -0.036 \\
         %\cite{sato2003unconventional} & 3.4  &2.02  & -0.014$^{\dagger}$ &0  & 0.028  & -0.0007(6)\\
         \cite{petit2007spinphonon} &2.3  & /  & -0.025 & -0.016 & 0.33 & 0.0008 \\

    \bottomrule
    \bottomrule
    \end{tabular}
    \caption{Hamiltonian parameters fitted to inelastic neutron scattering data.  \textsuperscript{$\dagger$} Note that in Ref. \cite{sato2003unconventional} only the $J_1^z-J_2^z$ difference was extracted.}
    \label{tabsupp:INSHam_params}
\end{table}

\begin{figure}[H]
    \centering
    \includegraphics[width=0.45\linewidth]{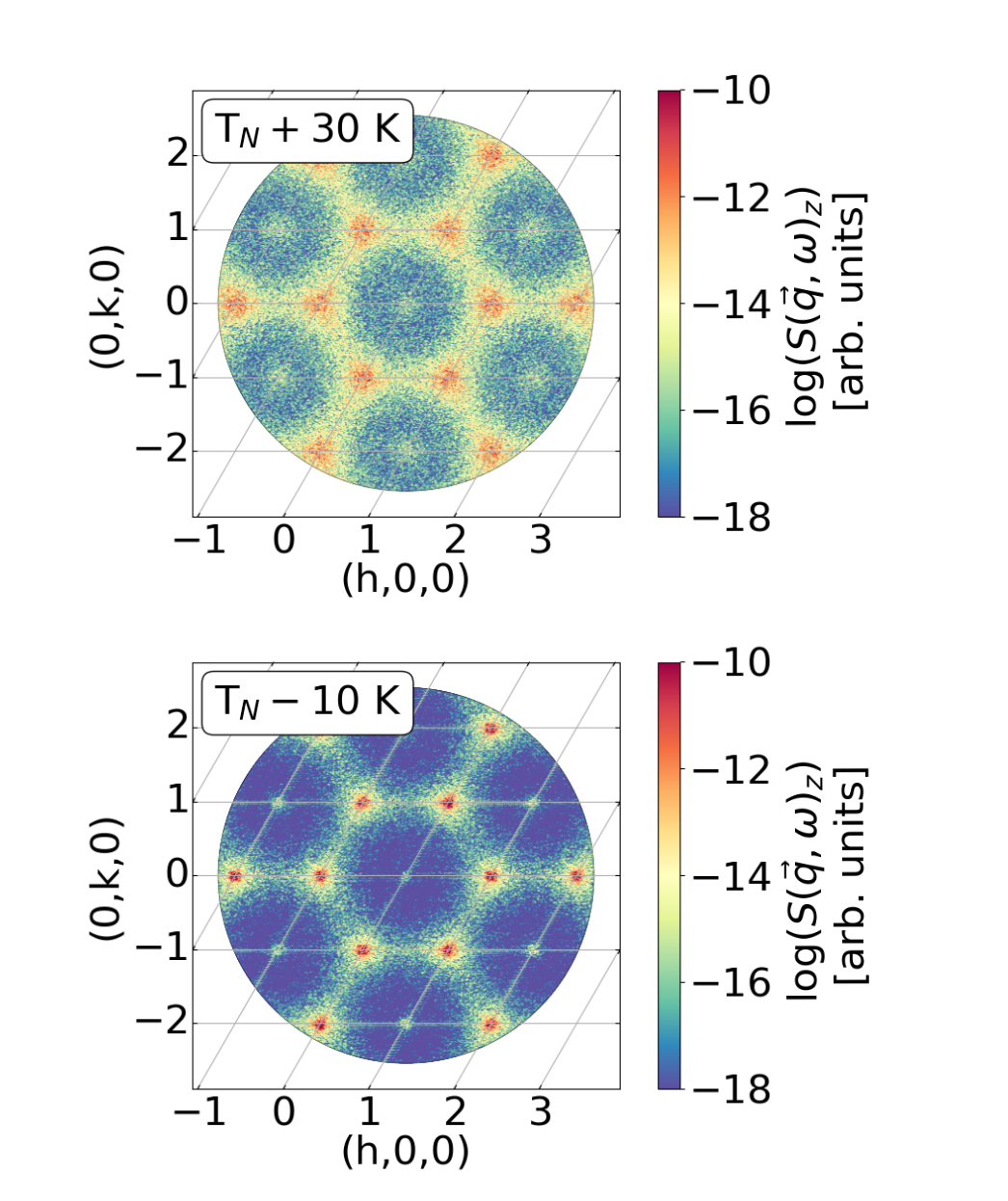}
    \caption{Calculated $S(\vec{q},\omega)_z$ at a constant energy $\hbar\omega=1.04$\,meV for the simple model Hamiltonian with only AFM coupling on a triangular lattice: $ \mathcal{H} =J_{1}^{(ab)}\sum_{\substack{\left<i,j\right>\\\text{1\textsuperscript{st}n.n}}}\hat{e}_i\cdot\hat{e}_j
   + J_{2} ^{(ab)}\sum_{\substack{\left<i,j\right>\\\text{2\textsuperscript{nd}n.n}}}\hat{e}_i\cdot\hat{e}_j $.}
    \label{TFL_sqwab}
\end{figure}

\begin{figure}[H]
    \centering
    \includegraphics[width=0.45\linewidth]{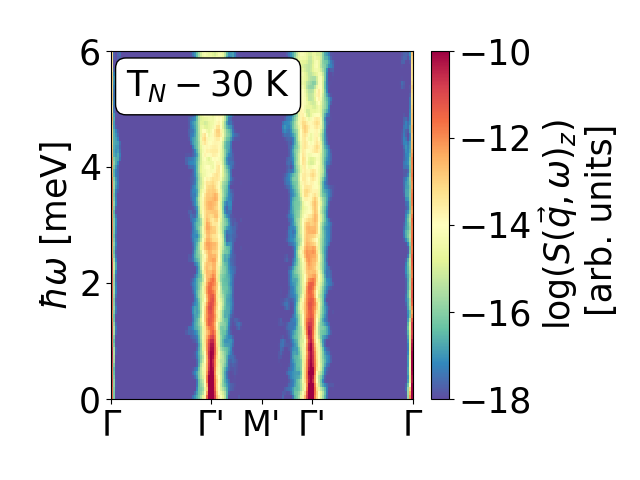}
    \caption{Calculated $S(\vec{q},\omega)_z$ for the following simple Hamiltonian model with only AFM coupling on a triangular lattice:   $\mathcal{H} =J_{1}^{(ab)}\sum_{\substack{\left<i,j\right>\\\text{1\textsuperscript{st}n.n}}}\hat{e}_i\cdot\hat{e}_j
   +J_{2}^{(ab)}\sum_{\substack{\left<i,j\right>\\\text{2\textsuperscript{nd}n.n}}}\hat{e}_i\cdot\hat{e}_j$. There is no energy gap at the $\Gamma'$ points due to the absence of $A_{zz}$.}
    \label{figsupp:sqwz_TFLtrim}
\end{figure}

\begin{figure}[H]
    \centering
    \includegraphics[width=1\linewidth]{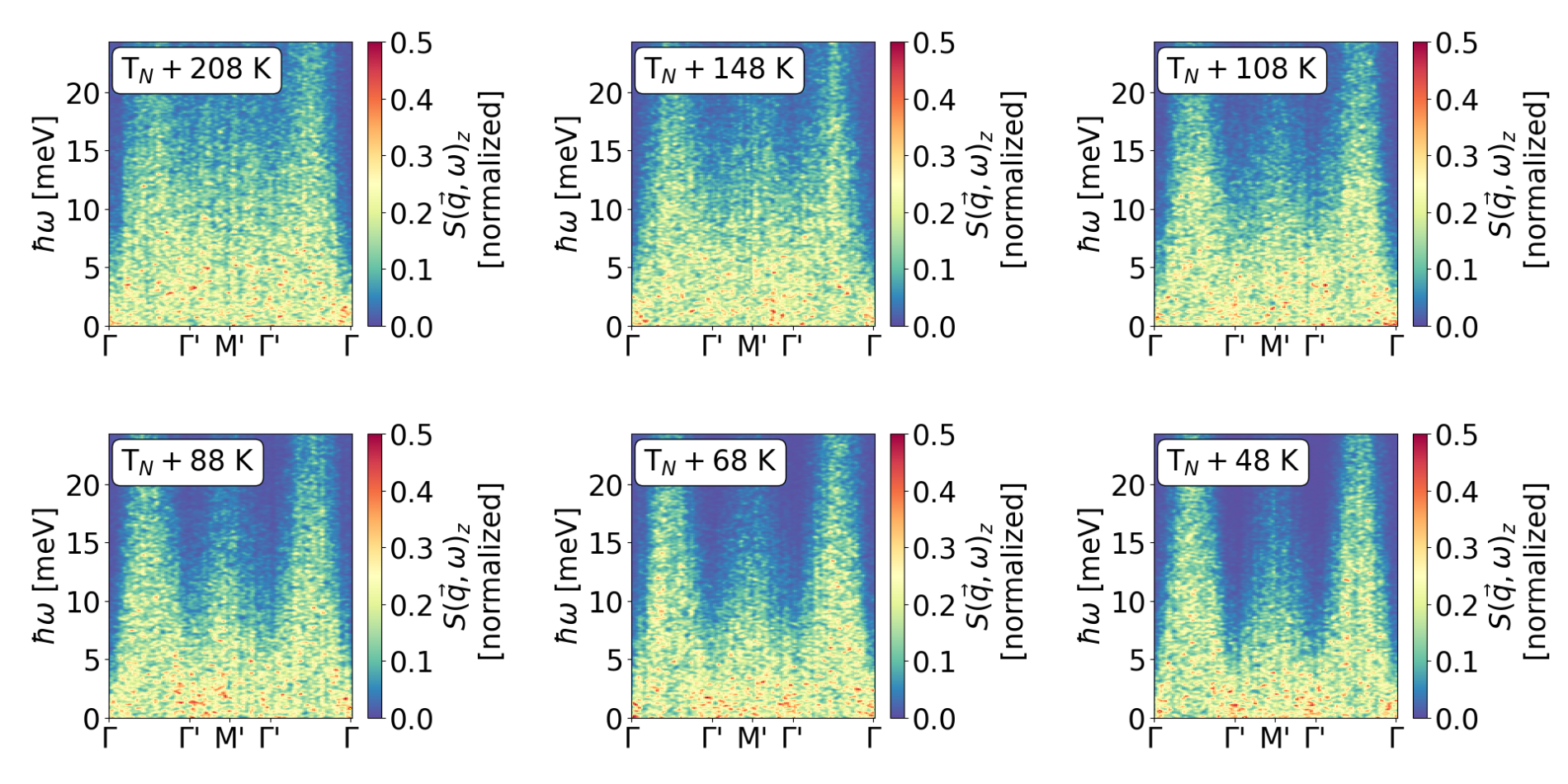}
    \caption{Calculated $S(\vec{q},\omega)_z$, normalized along each $\vec{q}$, at six temperatures above T\textsubscript{N}. We see an excitation continuum at all temperatures, which starts being bounded at high energies by the low-temperature magnon branches at $\simeq 150$\,K above T\textsubscript{N}.}
    \label{figsupp:sqwz_hightemp}
\end{figure}

\begin{figure}[H]
    \centering
    \includegraphics[width=1\linewidth]{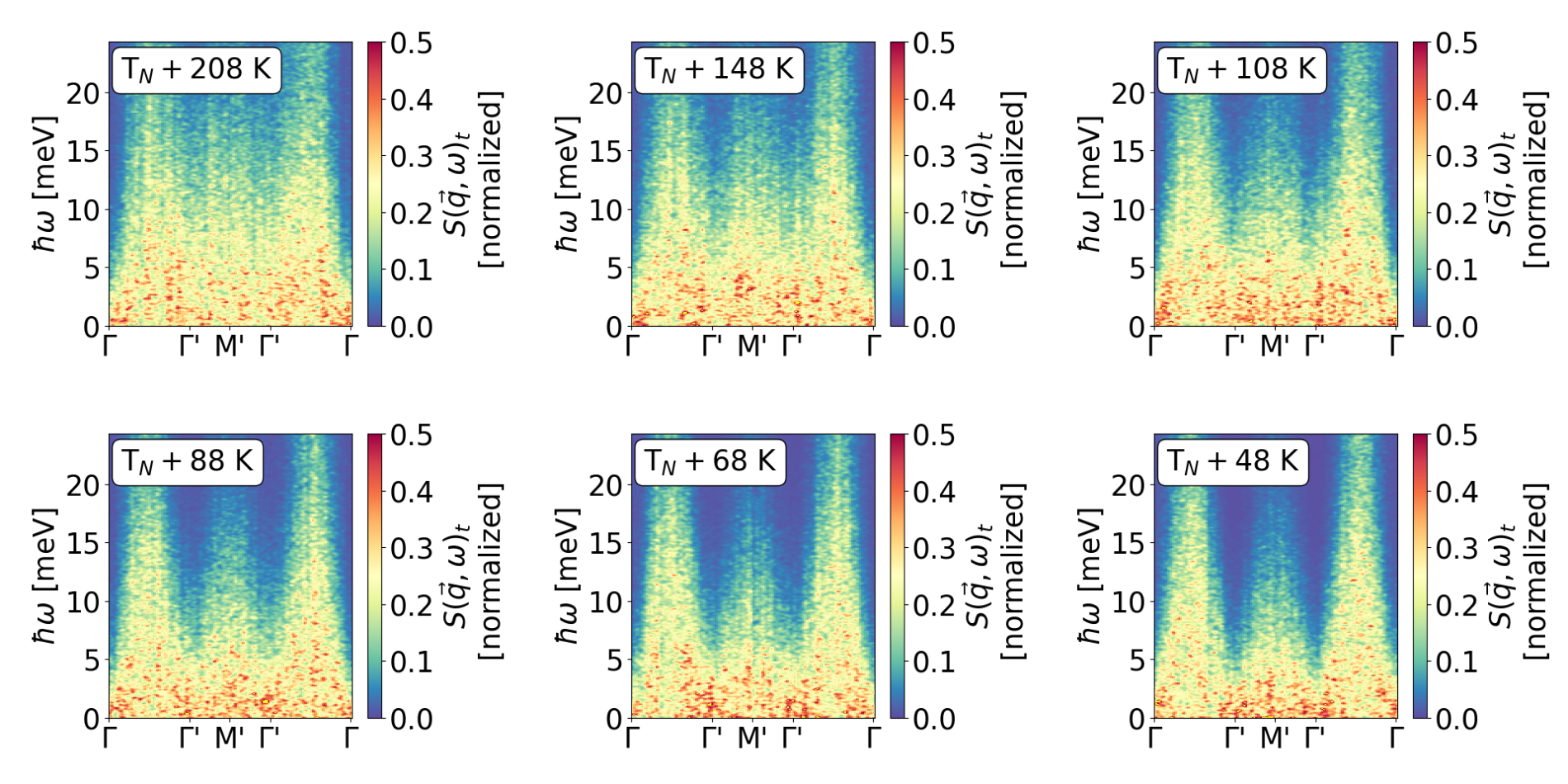}
    \caption{Calculated $S(\vec{q},\omega)_t$, normalized along each $\vec{q}$, at six temperatures above T\textsubscript{N}. We see an excitation continuum at all temperatures, which starts being bounded at high energies by the low-temperature magnon branches at $\simeq 150$\,K above T\textsubscript{N}.}
    \label{figsupp:sqwt_hightemp}
\end{figure}

\begin{figure}[H]
    \centering
    \includegraphics[width=0.9\linewidth]{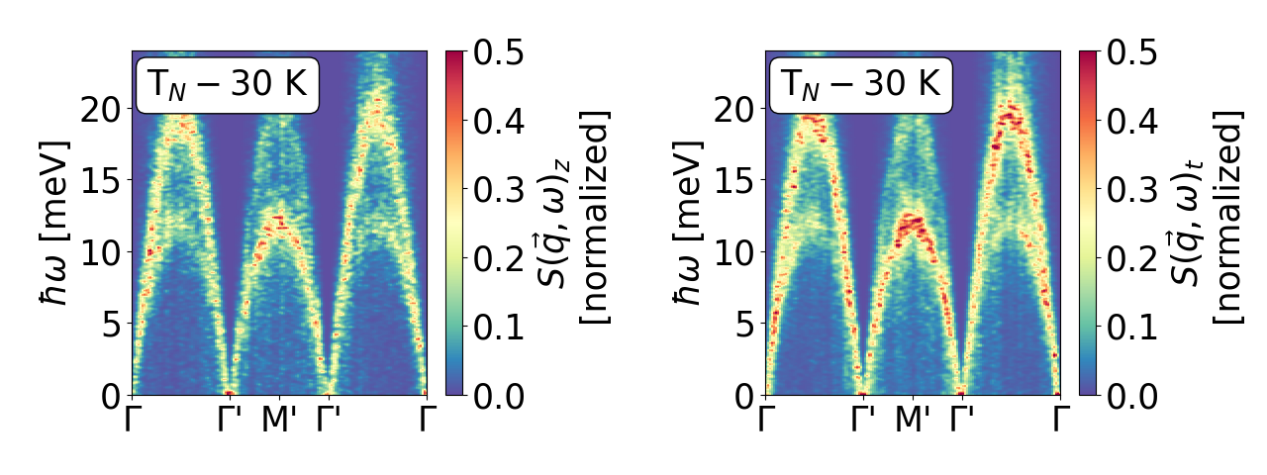}
    \caption{Calculated $S(\vec{q},\omega)_z$ and $S(\vec{q},\omega)_t$, normalized along each $\vec{q}$, for the following simple Hamiltonian model  with only AFM coupling on a triangular lattice:   $\mathcal{H} =J_{1}^{(ab)}\sum_{\substack{\left<i,j\right>\\\text{1\textsuperscript{st}n.n}}}\hat{e}_i\cdot\hat{e}_j
   +J_{2}^{(ab)}\sum_{\substack{\left<i,j\right>\\\text{2\textsuperscript{nd}n.n}}}\hat{e}_i\cdot\hat{e}_j$. Both the $z$ and $t$ spectra have both magnon branches and along all the paths in $\vec{q}$-space.}
    \label{figsupp:Sqw_TFL}
\end{figure}